\documentclass[aps,preprint, double-spaced,showpacserscriptaddress,floatfix,nofootinbib,reprint]{revtex4-1}
\usepackage{graphicx}
\usepackage{amssymb}
\usepackage{amsmath}
\usepackage{color}
\usepackage{eurosym}
\usepackage[yyyymmdd,hhmmss]{datetime}
\usepackage{MnSymbol}
\usepackage[english]{babel}
\usepackage{manfnt}
\usepackage{rotating}
\begin{document}
\title{Designing toroidal cavities for quantum computation} 
\author{C. A. L\"utken}
\thanks{lutken@gmail.com}
\affiliation{Dept.\;of Phys., University of Oslo, Norway}

\begin{abstract}
Toroidal microwave cavities are investigated for potential use in 
quantum information storage and computation. Since exact analytical 
results are not available for this geometry, extensive numerical simulation 
has been used to develop a universal phenomenological model 
(``spectral flow diagram").
This model is needed to guide the non-trivial design of toroidal resonators. 
A host of new modes that do not exist in cylindrical cavities are classified,
including novel counter-intuitive ground states, and ``dark nodal modes" 
that are decoupled from the environment in the absence of antennae. 
Numerical results are found to be in good agreement with experimental data. 
The existence of dark nodal modes in a shallow smooth cavity 
geometry that offers easy access for high quality surface treatment, suggests that 
high-Q toroidal cavities may be exploited for long-term storage of quantum 
information used in quantum processors. 
\end{abstract}
\maketitle

\section{Cavity geometry}

Consider a real or artificial atom in a two- or three-dimensional cavity.
The conventional way to view this is as a well-localised non-linear
system (the ``atom") perturbed by weak interactions with a non-local 
environment of linear oscillators (the radiation field, or photons).

Strongly coupled systems in cavity- and circuit-QED now available 
for experimental investigation challenges this conceptualisation, and
leads to an equally valid complementary picture where the radiation field
is being perturbed by the atom. This ``conceptual castling" has ramifications
for quantum computation, some aspects of which will be investigated here.

Our desire to mitigate decoherence leads to several 3D cavity design considerations,
including \emph{geometry, topology, manufacturing, decoupling, error correction, 
and scalability}.

Since exact analytical results are not available for tori, instead a taxonomy of 
modes for various cavities is compiled, derived from extensive numerical 
simulations using COMSOL. Because microwave textbooks usually only 
consider exceptionally simple geometries, like waveguides with rectangular 
cross-sections, this produces some surprises.

Curved boundaries introduce non-intuitive curvature 
corrections that sometimes can invalidate simplistic expectations. It is, for example, 
not always true that increasing the mode volume lowers the frequency of a mode, 
as may be expected from cuboid and cylindrical results familiar from waveguide engineering \cite{Pozar}. 
The frequency $f_{100}^{\rm TE} = c/(2 a)$ of the ground state 
${\rm TE}_{100}$ of a cuboid cavity with sides $b, c < a$ clearly decreases when 
the volume $V = a b c$ of the cavity grows by increasing $0 < a < \infty$. 
Surprisingly, the lowest mode frequency of a toroidal cavity with 
major and minor radii $R$ and $r$,\emph{ increases} when the volume 
$V = 2 \pi^2 r^2 R$ of the cavity grows by increasing the major radius 
$R$ (cf.\;Fig.\;\ref{fig:TorSpecFlow}).

Furthermore, curved cavities may support ``dark modes" (DM), whose electric fields 
by definition are completely decoupled from the environment in the absence of 
antennae, and may therefore have exceptionally high internal Q-values when 
the aluminium or niobium container is superconducting.

\section{Universal mode functions}

It is very useful to map all spectral data (numerical and experimental) onto dimensionless universal mode functions (spectral flow diagrams) that depend only on the chosen cavity geometry. Since (genus one) toroidal cavities (for which exact analytical solutions are not available) will be compared and contrasted with simpler (genus zero) cavities (for which analytical solutions are available), consider first cylindrical cavities. Comparison of toroidal and cylindrical modes gives considerable insight into the structure of toroidal modes, and allows us to devise a simple and rational classification scheme for toroidal modes.

These families of closed cavities have free parameters $p$ and $q$ that determine the shape and size of the cavity, most conveniently parametrised by the dimensionless aspect ratio 
$\epsilon = p/q\rightarrow\epsilon \ll 1$ when $p\ll q$. 
For genus zero cavities (cuboid, cylindrical, ellipsoidal,...)
the aspect ratio can take any positive real value ($0 < \epsilon < \infty$).

However, for genus one toroidal cavities, which is a product of two circles with minor 
and major radii $r$ and $R$, the aspect ratio $\epsilon = r/R$ cannot exceed 
one ($0 < \epsilon < 1$). Sometimes the inverse $1 < 1/\epsilon = R/r < \infty$ 
is called the aspect ratio of a torus, but $\epsilon < 1$ is a more convenient 
expansion parameter in perturbative analytical approximations.

Universal mode functions are obtained by plotting dimensionless ``frequencies" 
$F_{knm}(\epsilon) = f_{knm}(\epsilon)\times (d/c)$ as functions of the aspect ratio, 
where $f_{knm}$ are the resonance frequencies, $d = 2 r$ is the (minor) diameter 
of the (toroidal) cylindrical cavity, and $c$ is the speed of light in the dielectric filling 
the cavity (usually air or vacuum).

\begin{figure}[] 
\includegraphics[scale=0.19]{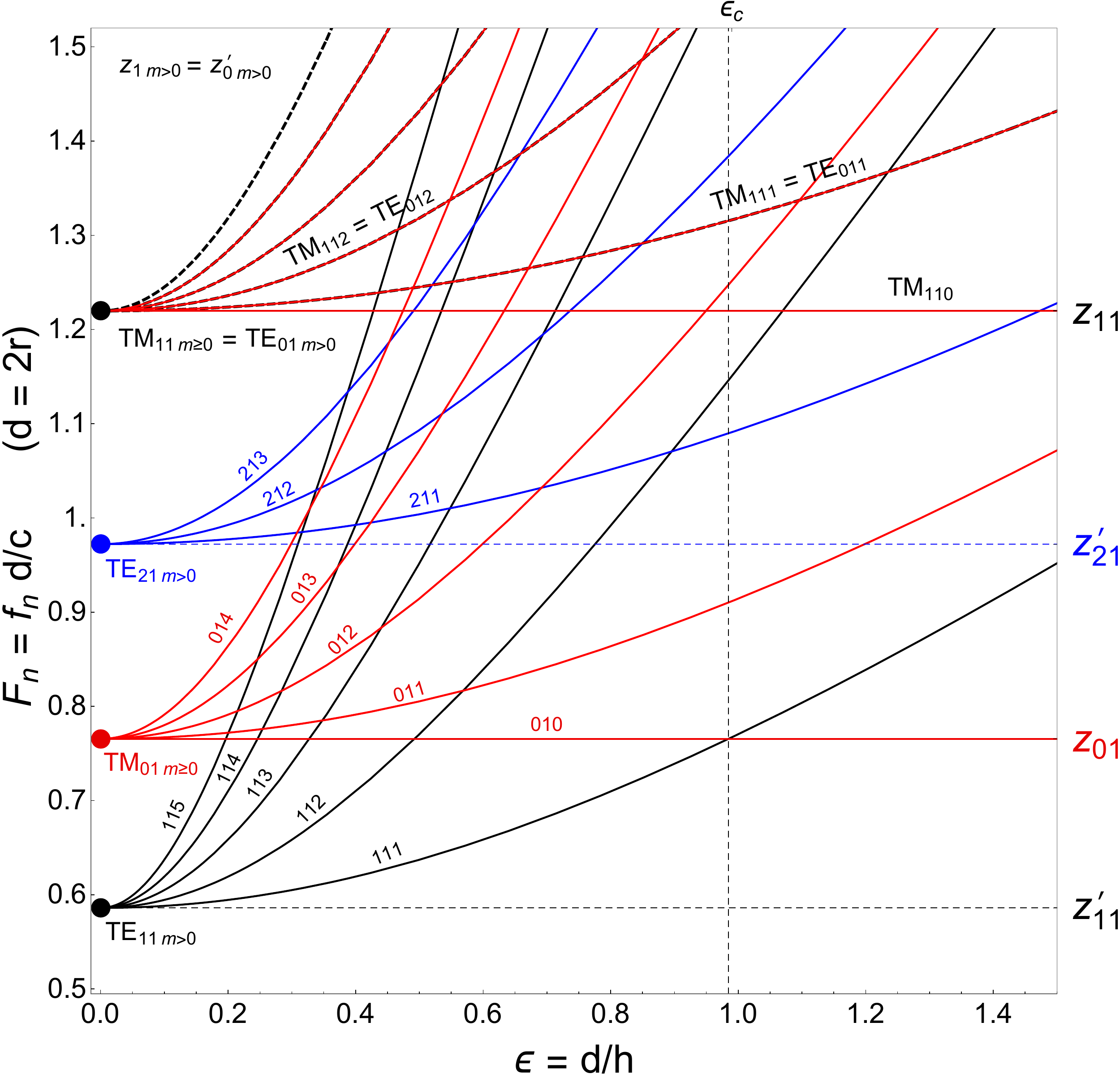}
\caption[Universal cylindrical spectral flow]
{Frequencies of dimensionless universal mode functions for any cylindrical 
closed cavity with diameter $d$ and height $h$. Curves are exact analytical solutions, 
and $c$ is the speed of light inside the cavity (usually empty or filled with air).
When $\epsilon\rightarrow 0$ the ``fans" degenerate to $m$-independent asymptotic values 
$z_{k1} = p_{k1}/\pi\;(k\geq 0)$ and $z^\prime_{k1} = p^\prime_{k1}/\pi\;(k\geq 1)$, 
where $p_{k1}$ is the first zero of the Bessel function $J_k$, and $p^\prime_{k1}$ 
is the first zero of $J_k^\prime$. The vertical dashed line at $\epsilon_c\approx 0.99$
shows where the ground state changes from ${\rm TE}_{111}$ to ${\rm TM}_{010}$.}
\label{fig:CylSpecFlow}
\end{figure}

\subsection{Cylindrical resonators}

\begin{figure*}[] 
\includegraphics[scale=0.49]{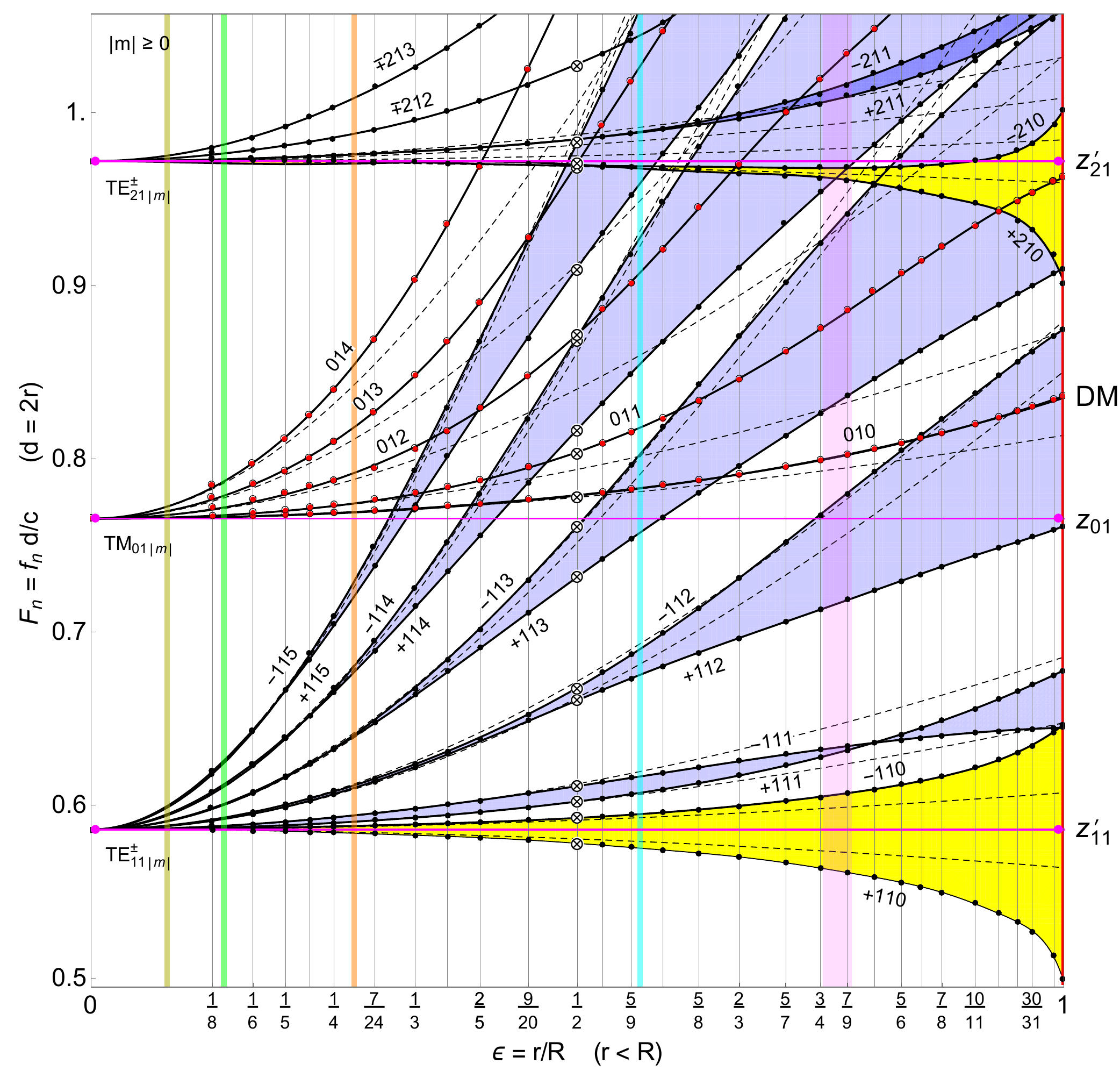}
\caption[Universal toroidal spectral flow]
{Dimensionless universal spectral flow diagram for any toroidal cavity.
The 667 solid plot markers are data points from numerical simulations, 
solid black curves are fits to these data, dashed curves are analytical 
solutions at $\mathcal O(\epsilon^2)$  (Eqs.\;\ref{eq:TE}\,-\,\ref{eq:TM}).
The speed of light inside the cavity (usually empty or filled with air) is $c$, 
$p_{k 1} = \pi z_{k 1} \; (k \geq 0)$ is the first zero of the Bessel function $J_k$,
and $p_{k 1}^\prime = \pi z^\prime_{k 1}  \; (k > 0)$ is the first zero of $J_k^\prime$.
Crossed plot-markers ($\otimes$) are dimensionless experimental frequencies 
$F_n^x$ of the $R = 2r = 20\,{\rm mm}$ cavity, obtained with a VNA at room 
temperature (cf.\;Fig.\;\ref{fig:AluModeChart}). The dark mode ${\rm DM = TM}_{010}$ was found using an unconventional antenna. 
TE modes split into pairs of eigenstates
${\rm TE}^\pm_{knm} = \pm\vert k, n, m\rangle = \pm knm$ with positive and 
negative parity $P = \pm 1$.}
\label{fig:TorSpecFlow}
\end{figure*}

Fig.\;\ref{fig:CylSpecFlow} shows the resulting spectral flow derived from 
exact cylindrical mode function that are given by zeros of Bessel functions
 \cite{Pozar},
\begin{eqnarray}
\left(F_{knm}^{\rm TE}\right)^2 &=& \left[f_{knm}^{\rm TE}(\epsilon) \,d/c\right]^2 
= {z^{\prime\;2}_{k n} + m^2  \left(\frac{\epsilon}{2}\right)^2},\label{eq:cyl1} \\
\left(F_{knm}^{\rm TM}\right)^2 &=& \left[f_{knm}^{\rm TM}(\epsilon)\,d/c\right]^2 
= {z_{k n}^2 + m^2  \left(\frac{\epsilon}{2}\right)^2}, \label{eq:cyl2}
\end{eqnarray}
where $n\geq 1$, $\epsilon = d/h$, and $h$ is the height of the cylinder.
These modes degenerate to $m$-independent points $z_{k n} = p_{k n}/\pi$ 
and $z_{k n}^\prime = p_{k n}^\prime/\pi$ when $\epsilon\rightarrow 0$, 
where $p_{k n}$ is the n'th zero of the Bessel function $J_k$,
and $p_{k n}^\prime$ is the n'th zero of $J_k^\prime$.
Each of these points sprouts a fan of spectral curves that mix in a 
confusing manner for finite values of $\epsilon$, which can only be 
disentangled by a proper classification of the modes and their asymptotic 
values, as shown in this diagram. A similar, but more interesting and 
complicated spectrum obtains for tori (cf.\;Fig.\;\ref{fig:TorSpecFlow}). 

Notice that $z_{1n} = z_{0n}^\prime$, so TE-modes must have $k > 0$ 
since TM-modes cannot be degenerate with TE-modes. Furthermore, 
while cylindrical TE-modes must have $m\,\neq\,0$, this is not the case for tori,
whose ground states therefore are qualitatively different from cylindrical ground 
states. Every possible cylindrical spectrum is encoded in this diagram.

\begin{figure}[b] 
\includegraphics[scale=0.27]{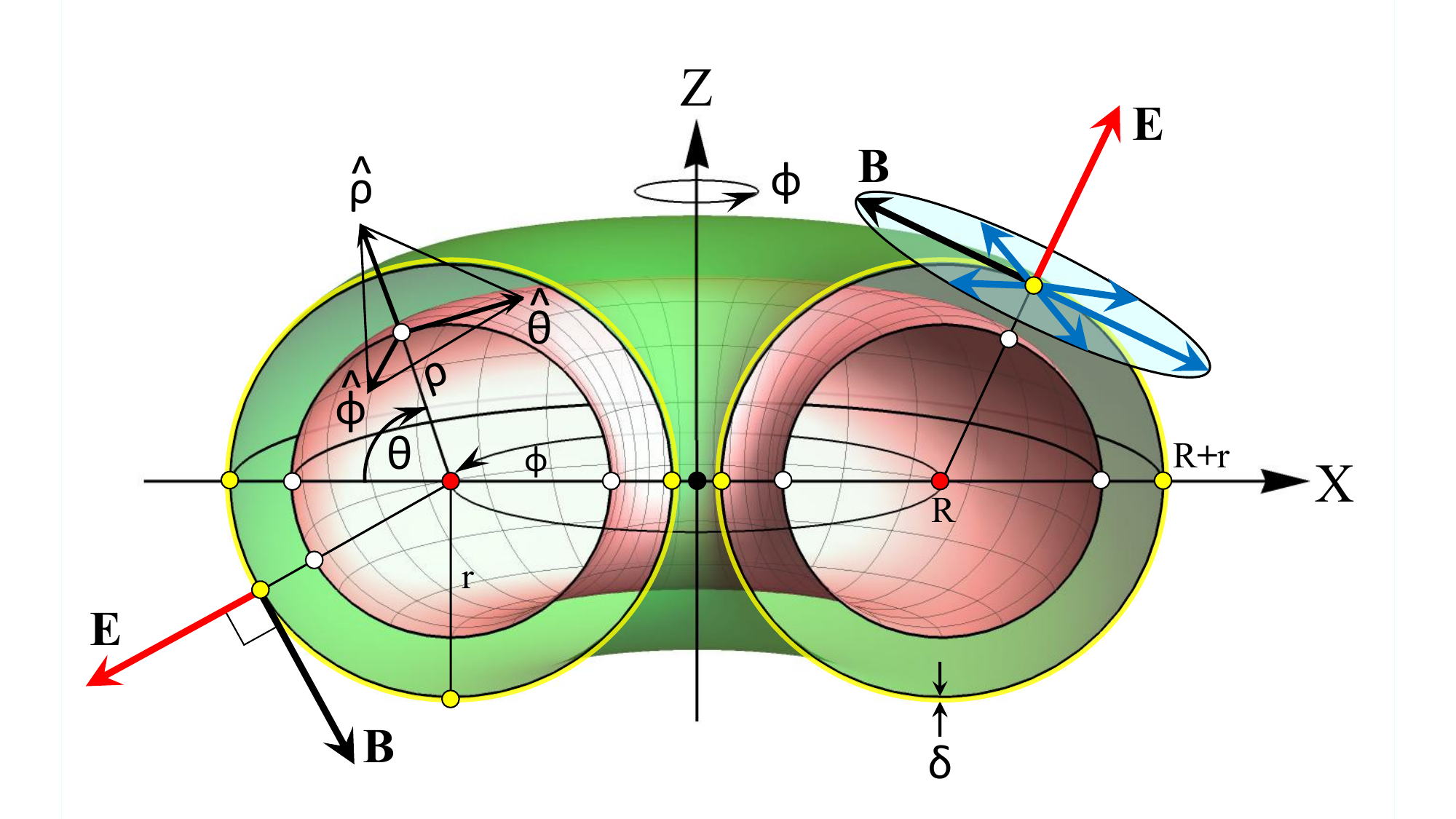}
\caption[Toroidal-poloidal coordinates]
{Toroidal-poloidal coordinates $(\rho,\phi,\theta)$ used in spectral analysis 
of microwaves in an evacuated toroidal cavity.}
\label{fig:PolarCoord}
\end{figure}

\subsection{Toroidal resonators}

Fig.\;\ref{fig:PolarCoord} shows the doubly-periodic polar (toroidal-poloidal) coordinates $(\rho,\phi,\theta)$ used here in the analysis of electromagnetic modes 
inside an otherwise empty toroidal cavity inside a lump of highly conductive 
metal, e.g., aluminium that is superconducting below about $1\,K$.
Also shown is the metallic boundary conditions for the electromagnetic field 
at the surface of the cavity (green).
In normal lossy conductors (superconductors) there is a very small exponentially damped residual surface resistance $R_s \propto \exp(-2\rho/\delta)$, 
where $\delta$ is the skin depth (penetration depth) of the cavity wall (yellow).

 Similar to cylindrical cavities, symmetries and boundary conditions constrain
 the eigenmodes (resonances) to a discrete spectrum labeled by three integer 
 mode numbers.
Symmetry in the \emph{poloidal} angle (elevation) $\theta$ only admits 
eigenmodes labelled by an integer $k\geq 1$ ($k\geq 0$) for TE (TM) 
modes. Electromagnetic boundary conditions on the inner wall of 
a toroidal cavity in a highly conductive material gives a second 
mode number $n\geq 1$.
Finally, symmetry in the \emph{toroidal} (azimuthal) angle $\phi$ only admits 
eigenmodes labelled by an integer $\vert m\vert \geq 0$.

The numerical results in Fig.\;\ref{fig:TorSpecFlow} shows that, unlike the cylindrical case, toroidal TE modes all split into pairs with different energy, labelled by a new mode number here called \emph{parity}.
The parity $P=\pm 1$ of a TE resonance mode in a toroidal cavity 
is by definition positive (negative) if the mode is even (odd) in the poloidal 
angle $\theta$, while TM modes are not parity eigenmodes \cite{Fred:2023}.
In short, toroidal TE modes are labelled by four integers,  
${\rm TE}^P_{k, n, m} = P \vert k, n, m\rangle$, while toroidal TM modes
are labelled by three integers, ${\rm TM}_{k, n, m} = \vert k, n, m\rangle$.
In the diagrams used here the kets are not needed and therefore omitted.
This labelling is complete, since there are no TEM modes in toroidal cavities
 \cite{Fred:2023}.

Unfortunately, exact expressions for eigenfrequencies in tori are not available,
but it is possible to find approximate expressions by expanding to lowest order 
in $\epsilon = r/R \ll 1$ ($r\ll R$). Notice first that the numerical simulation 
shown in Fig.\;\ref{fig:TorSpecFlow} degenerates to the exact cylindrical result in
Fig.\;\ref{fig:CylSpecFlow} in the limit $\epsilon\rightarrow 0$, so the
leading constant expansion coefficients are the Bessel zeros $z$ and $z^\prime$.
Since a large skinny torus can be constructed by bending a long thin
cylinder of radius $r$ and height $h$ into a torus with minor and major 
radii $r$ and $R$, the toroidal expansion parameter 
$\epsilon/(2\pi) = r/(2\pi R)$ should correspond to the cylindrical parameter 
$\epsilon/2 = r/h$ with $h \approx 2\pi R$ [cf.\;Eq.\;(\ref{eq:cyl1})].
A reasonable ansatz is therefore that the toroidal frequencies take the form
\begin{eqnarray}
{\rm TE^\pm:}\quad \left(F^{\pm\rm{TE}}_{knm}\right)^2 &=&  
{z^{\prime\;2}_{kn} + \alpha^{\pm}_m \left(\frac{\epsilon}{\pi}\right)^2} 
+ \mathcal{O}(\epsilon^4),\label{eq:TE}\\
{\rm TM:}\quad \left(F^{\rm TM}_{knm}\right)^2 &=& 
{z_{kn}^2 +  \alpha_m \left(\frac{\epsilon}{\pi}\right)^2} 
+ \mathcal{O}(\epsilon^4).\label{eq:TM}
\end{eqnarray}
The expansion coefficients $\alpha$ and $\alpha^P$ are not expected to be 
cylindrical ($\propto m^2$), because both the curvature and boundary 
conditions of the torus are different from a cylinder. 
Perhaps the easiest way to determine the coefficients 
is to fit polynomials in $m^2$ to the numerical results, but to lowest order
perturbative analytical results are available, $\alpha^P_m = m^2 - P/4$ and 
$\alpha_m = m^2 + 3/4$ \cite{Fred:2023}. 

Fig.\;\ref{fig:TorSpecFlow} shows the spectral flow functions derived from extensive 
numerical simulations of toroidal cavities. Crossed plot-markers ($\otimes$) are 
experimental frequencies for the toroidal aluminium cavity with 
$R = 2r = 20 \,{\rm mm}$ ($\epsilon = r/R = 1/2$, cf.\;Fig.\;\ref{fig:ProtoNodal}, left), 
shifted by one very small systematic error. If the minor radius of the cavity 
was machined larger than the nominal value $r = 10\,{\rm mm}$ by only 
$\delta r\approx 25\, \mu\rm{m}$,
this shifts frequencies on average -22\,MHz, in which case these numerical and 
experimental data differ on average by about 2\,MHz (and at most by about 10\,MHz).
The experimental data, discussed in more detail in Fig.\;\ref{fig:AluModeChart}, 
completely eclipse the numerical results. 
As more toroidal cavities with different aspect ratios are manufactured, 
this diagram will be populated with more experimental data that presumably 
will further validate both the numerical and analytical work presented here.

To illustrate the universality of this diagram, consider some well-known tori, 
represented here by vertical coloured bands.
Pink dots on the left is the dimensionless spectrum in a microwave cavity the size 
of the Large Hadron Collider at CERN
($\epsilon\approx 1.9\,{\rm m}/ 4.47\, {\rm km} = 0.00045\dots \approx 0$),
given by zeros of the Bessel functions $J_k$ and its derivative $J_k^\prime$.
The dark green band on the far left approximates the aspect ratio 
$\epsilon\approx 65 \,{\rm m}/830\, {\rm m} \approx 0.08$ of the proposed 
rotating NASA-Stanford toroidal space station, designed to house 10 000 people.
The green band on the left indicates typical aspect ratios 
$\epsilon \approx 1.4\, {\rm mm}/12 \,{\rm mm} \approx 0.12$ 
for wedding bands (small skinny tori filled with gold). 
The orange band on the left approximates the aspect ratio 
$\epsilon\approx 23 \,{\rm cm}/85 \,{\rm cm} \approx 0.27$ of a CAT scanner.
The blue band in the middle approximates the aspect ratio 
$\epsilon\approx 3.5 \,{\rm m}/6.2 \,{\rm m} \approx 0.56$ of the ITER Tokamak 
in the south of France, which will be the world's largest toroidal fusion reactor. 
The pink band on the right spans aspect ratios around
$\epsilon\approx 1.7 \,{\rm cm}/2.2\,{\rm cm} \approx 0.77$ for bagels and donuts.

The torus with aspect ratio $\epsilon = 1$, indicated by the vertical red line on 
the right, is sometimes called a ``horn torus", but a more precise name is
 \emph{nodal cavity}, derived from the mathematically precise term 
 ``nodal elliptic curve". 
For the numerical algorithm to work the simulation was done at a 
slightly smaller value $\epsilon = 10 \,{\rm mm}/10.01 \,{\rm mm} \approx 0.999$.
This is also more realistic as an exact nodal torus with $\epsilon = 1$ 
cannot be manufactured on a CNC. If the ``thread" in the middle of the 
cavity is snipped, then it is no longer a torus (genus one), 
but a topologically distinct cavity (genus zero) that is homeomorphic to a sphere.

\begin{figure}[] 
\includegraphics[scale=0.78]{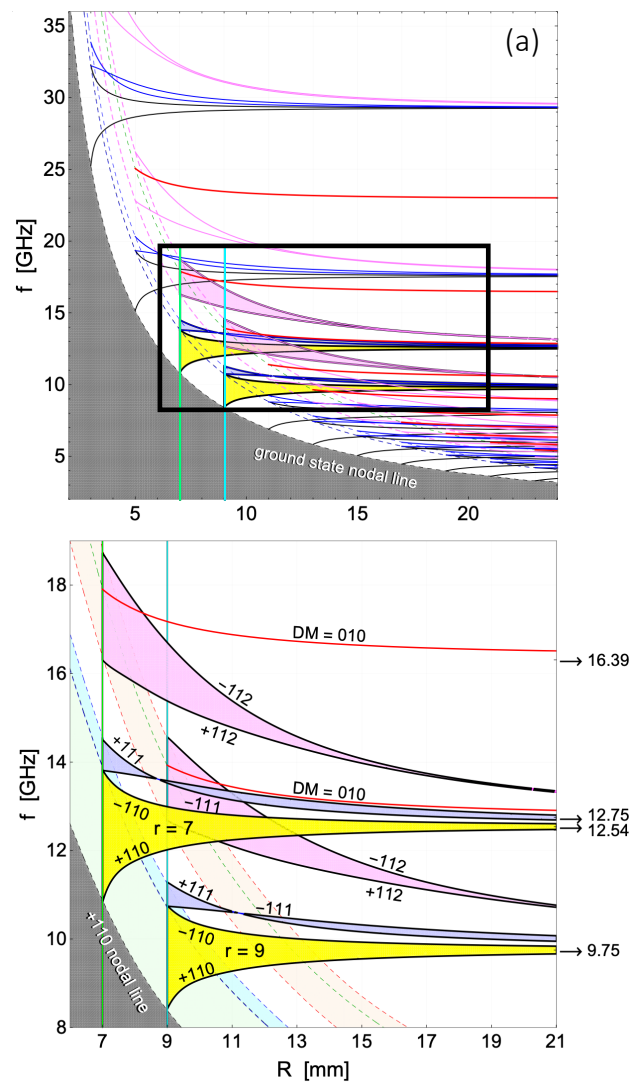}
\caption[Non-universal toroidal spectrum]
{Non-universal toroidal spectrum, showing the first seven frequencies (GHz) as a 
function of the two toroidal radii measured in milli-meters.
The region inside the black frame in (a) is magnified in (b). Electric modes with
opposite parity border coloured bands, while the magnetic ground state 
(DM = 010) is red.}
\label{fig:ModeSpectrum}
\end{figure}

Comparison with the cylindrical case (Fig.\;\ref{fig:CylSpecFlow}) reveals 
a number of important differences. Unlike cylindrical modes, transverse electric (TE) 
modes split into pairs of eigenmodes ${\rm TE}^\pm$ that have different energy 
and \emph{parity} $P=\pm 1$, which by definition means that the mode is even 
(odd) in the elevation (poloidal angle) $\theta$ (cf.\;Fig.\;\ref{fig:PolarCoord}).Transverse magnetic (TM) modes do not split.

The new ${\rm TE}^\pm_{k10}$ modes ($k \geq 1$) on the boundaries of the 
yellow regions do not exist in a cylinder, and are qualitatively different. 
The anomalous behaviour of ${\rm TE}^+_{k10}$ modes ($k \geq 1$; 
on the lower boundary of the yellow regions) is counter-intuitive since 
the frequency \emph{ increases} when the cavity volume 
$V=2 \pi^2 r^2 R$ grows by increasing $R$, at fixed $r$ 
\mbox{($0\leftarrow\epsilon$)}.

Metallic boundary conditions prohibits the electric field of ${\rm TM}_{k10}$ modes 
($k \geq 0$) from touching the cavity wall. These modes, which here will be called
 ``dark modes" (DM), should therefore have exceptionally high $Q$-values.
For $\epsilon \gtrsim 0.85$ the dark mode ${\rm TM}_{010}$ is always the sixth 
member of the spectrum. The dark nodal mode is of particular interest, 
since this geometry maximises the Q-factor, and also the gaps
$\delta F^\pm_{010} = (d/c)\, \delta f^\pm_{010}  = F_{\pm 112} - F_{010}$ 
to neighbouring modes.

For a nodal cavity with outer diameter $D = 4R = 80 \,{\rm mm}$
$c/d \approx 7.5\, {\rm GHz}$, and 
the gaps are $\delta f^+_{010} \approx -600 \,{\rm MHz}$ and 
$\delta f^-_{010} \approx +300\, {\rm MHz}$. For a quasi-nodal cavity with 
$R\approx r = 10\,{\rm mm}$  ($\epsilon \approx0.999$) the gaps are 
$\delta f^+_{010} \approx - 1.13 \,{\rm GHz}$ and 
$\delta f^-_{010} \approx + 0.58 \,{\rm GHz}$.
The splitting between the two lowest states ${\rm TE}^\mp_{110}$ is even larger, 
but in a nodal cavity the ground state ${\rm TE}^+_{110}$ has no electric field in 
almost all the cavity, and is therefore unfit for quantum computing.

Exact analytical solutions for a torus remain unknown, because Maxwell's 
equations are not separable in this case. This is somewhat surprising in view of 
the abundance of symmetry in a torus, but it is no longer a show-stopper since
the spectral problem, for practical purposes, has been solved by using 
powerful computers, numerical simulation (COMSOL), and an analytical 
calculator (MATHEMATICA), which until recently were not available. 
Previously attempts were made to circumvent the problem by inserting a 
conductive membrane that essentially cuts the torus into a curved cylinder. 
It is now clear that this is not a good idea since it misses $1/3$ of the 
eigenmodes available in an unadulterated torus.

\emph{In summary, every possible cylindrical spectrum is encoded in Fig.\;\ref{fig:CylSpecFlow},
and every possible toroidal spectrum is encoded in Fig.\;\ref{fig:TorSpecFlow}.}

Fig.\;\ref{fig:ModeSpectrum} is a decompressed non-universal toroidal mode chart, showing 
the first few frequencies (GHz) as a function of the two toroidal radii measured
in milli-meters. The inset shows spectral fans of modes as a function of the 
major radius R, for odd integer values $r = 3, 5, 7, \dots, 23\,{\rm mm}$ of the 
minor radius. It is derived from the interpolation curves in the universal flow diagram
(Fig.\;\ref{fig:TorSpecFlow}). For clarity Fig.\;\ref{fig:ModeSpectrum} (b) 
shows only the fans for 
$r = 7\,{\rm mm}$ and $r = 9\,{\rm mm}$. The dashed curves 
(``nodal lines") shows the nodal spectrum as a function of $r = R$.  
The gray region is forbidden because the minor radius of a torus cannot 
be larger than the major radius  ($r \leq R$). 
The Q-value (internal quality factor) of some low-frequency modes
can be read off Fig.\;\ref{fig:Qplot}.
This decompressed spectral chart may be useful for 
designing toroidal cavities with specific desired properties.

\section{Toroidal cavities}

\begin{figure}[] 
\vskip 0mm
\frame{\includegraphics[scale=0.26]{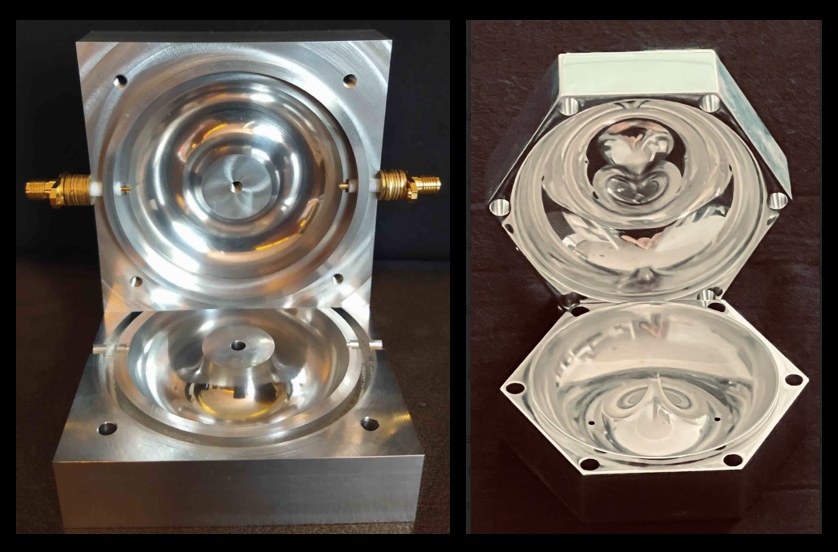}}
\caption[Prototypes]{ 
Left: Prototype toroidal cavity with $\epsilon\approx 0.5$ 
($R = 2r = 20 mm$). An experimental spectrum obtained with a VNA at room 
temperature is shown in Figs.\;\ref{fig:TorSpecFlow} and \ref{fig:AluModeChart}.
Right: Prototype quasi-nodal toroidal cavity with $\epsilon\approx 0.9$.}
\label{fig:ProtoNodal}
\end{figure}

\begin{figure*}[] 
\includegraphics[scale=1.25]{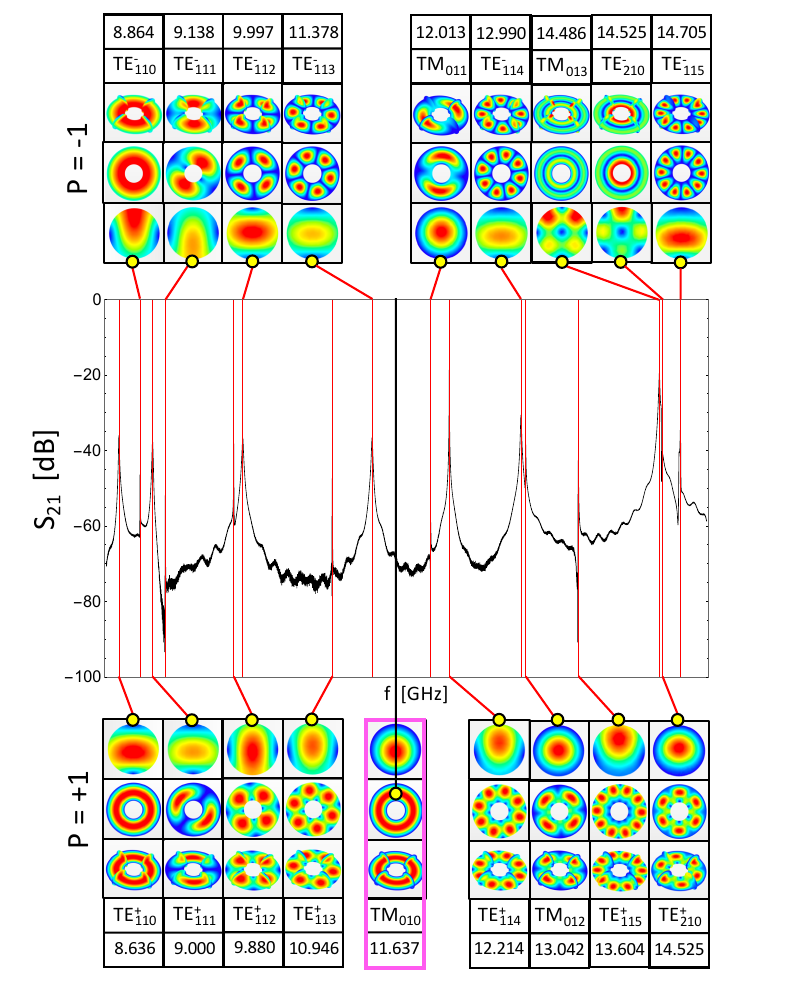}
\caption[Toroidal mode chart]
{Mode charts obtained by numerical simulation of the 
electric field density of the first 18 $\rm TE^+$ (bottom), $\rm TE^-$ (top), 
and $\rm TM$ modes in a toroidal cavity with 
major and minor radii $R = 2r = 20\;{\rm mm}$ ($\epsilon = r/R = 1/2$).
The number attached to each mode portrait is its frequency $f$ in GHz.
Middle: Experimental (transmission) spectrum $S_{21}$ of the cavity shown 
in Fig.\;\ref{fig:ProtoNodal} (left), obtained with a VNA and radial antennae.
This does not pick up the dark mode ($f_{\rm DM} = 11.637$ GHz), 
which was found using a bespoke ``azimuthal" antenna 
$-\kern-4pt-\kern-4pt($ oriented along the toroidal angle $\phi$. 
The spectral scan (black) is in very good agreement with the 
numerical simulation (vertical thin red lines). The experimental value of 
$f_{\rm DM}$ is also in good agreement with the numerical simulation 
($\otimes$ icon on the $\rm TM_{010}$ curve in Fig.\;\ref{fig:TorSpecFlow}).}
\label{fig:AluModeChart}
\end{figure*}

Two prototype toroidal cavities fabricated from 
aluminium are shown in Fig.\;\ref{fig:ProtoNodal}).
Fig.\;\ref{fig:AluModeChart} compares numerical simulations of the mode geometries
with the real VNA spectrum of the cavity machined in aluminium.
The numerical simulation gives frequencies ${\rm TE}^+_{110} = 8.662\,{\rm GHz}$, 
${\rm TE}^-_{110} = 8.884\,{\rm GHz}, \dots $. For comparison with real data 
the simulated spectrum has been shifted by $-22\;{\rm MHz}$ 
(almost invisible on this plot), 
corresponding to one very small systematic fabrication error in the minor radius 
($\delta r\approx 25\, \mu\rm{m}$).

The ground and first excited states in this cavity are approximately toroidal, 
but strongly coupled to the environment via the cavity wall.

Experimental resonance values (in GHz) are attached to the mode charts.
At this scale the real resonances, which are about four orders of magnitude 
larger than the background, are indistinguishable from the numerical simulation 
(Fig.\;\ref{fig:TorSpecFlow}). This confirms that the simulation is a reliable 
tool for spectral analysis. The magnetic ``ground state" 
\mbox{${\rm TM}_{010} = 11.637\,{\rm GHz}$} is not seen in this experiment, 
but  the frequency of this resonance was obtained separately using a custom made
antennae adapted to the task. As seen in Fig.\;\ref{fig:TorSpecFlow} it is also 
in agreement with the simulation.

\begin{figure}[] 
\includegraphics[scale=0.22]{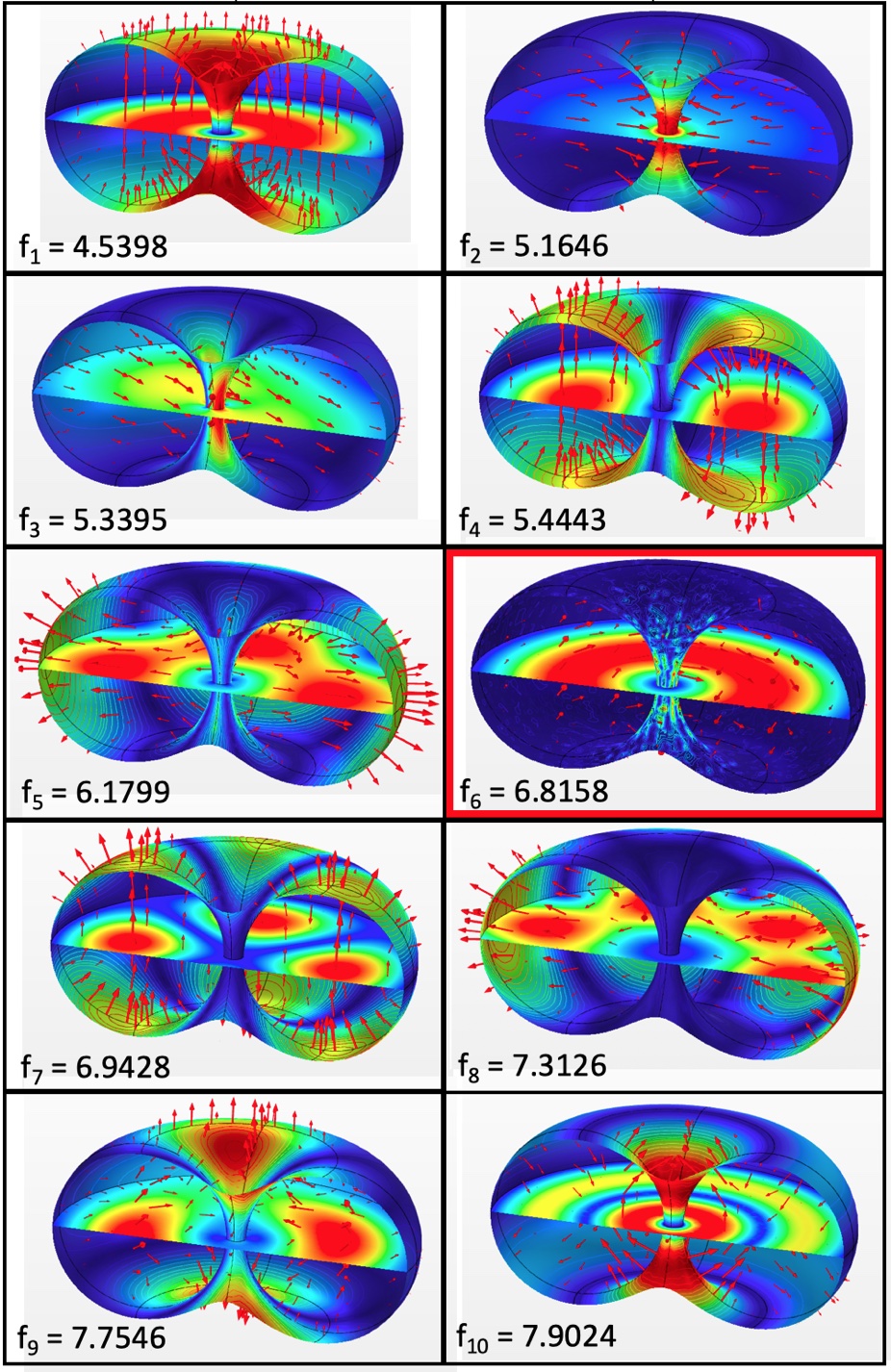}
\caption[Quasi-nodal modes]
{Electric field density of the first ten modes in a quasi-nodal cavity with 
$R = 20\,{\rm mm}$ and $r = 18\,{\rm mm}$ ($\epsilon = 0.9$).}
\label{fig:QuasiNodalModes}
\end{figure}

Observe that the electric field of the ninth mode
\mbox{${\rm TM}_{010} = 11.637\,{\rm GHz}$}
in Fig.\;\ref{fig:AluModeChart} (inside the purple frame) has support only on 
a torus nested inside the cavity with ($\epsilon = 0.5$), 
i.e., it does not touch the cavity walls. 
This ``magnetic ground state" is therefore a ``dark mode" (DM).

\begin{figure}[] 
\includegraphics[scale=0.06]{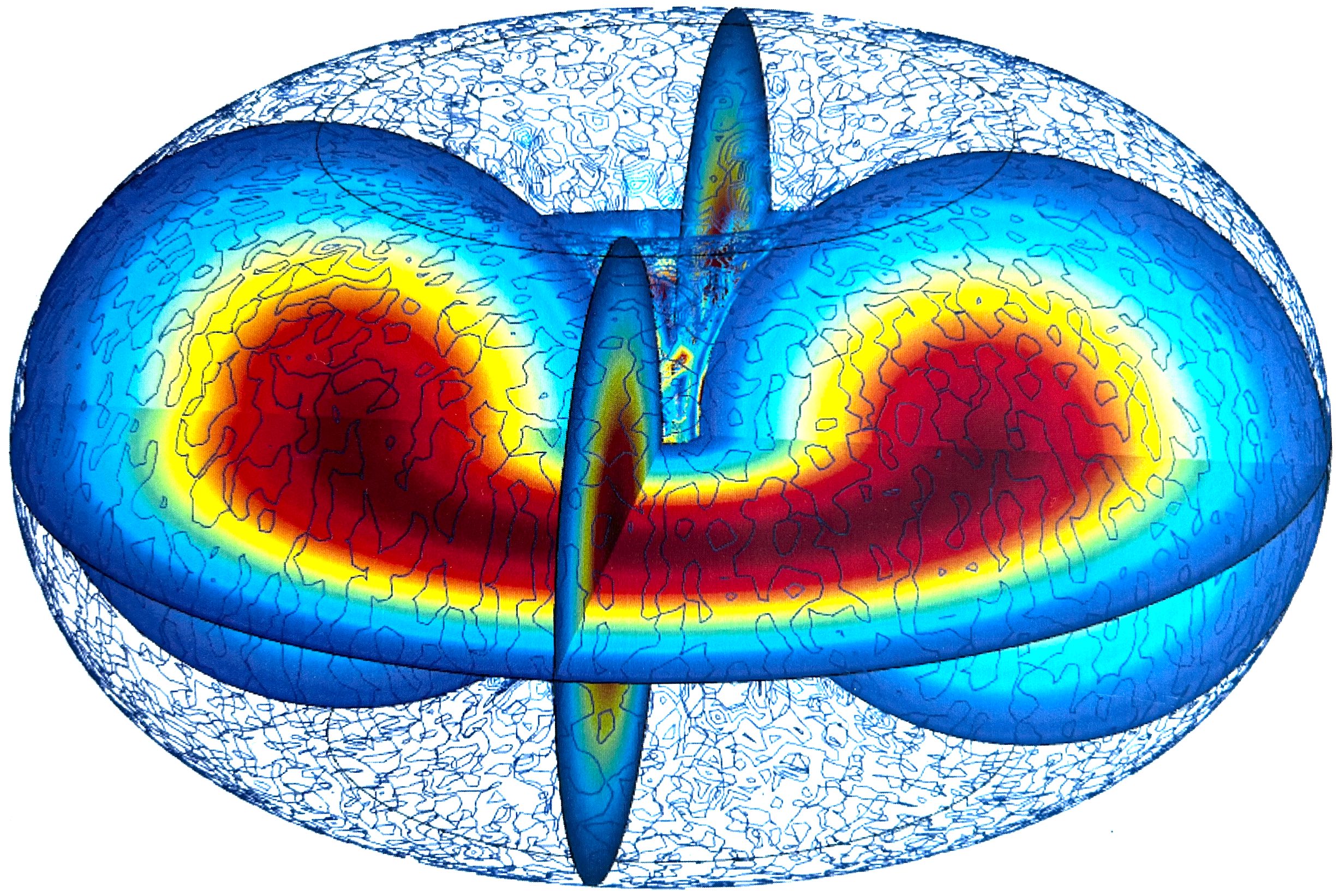}
\caption[Dark mode]
{A closer look at the dark quasi-nodal mode ${\rm TM}_{010} = 6.816\,{\rm GHz}$ 
shown inside the red frame in Fig.\;\ref{fig:QuasiNodalModes}.}
\label{fig:DarkTorMode}
\end{figure}

Fig.\;\ref{fig:QuasiNodalModes} shows the electric field density of the first 
ten modes in a quasi-nodal cavity with $R = 20\,{\rm mm}$ and $r = 18\,{\rm mm}$ 
($\epsilon = 0.9$). Fig.\;\ref{fig:DarkTorMode} shows the dark mode 
${\rm TM}_{010} = 6.816\,{\rm GHz}$ in more detail. 
The surface of the cavity is translucent, and the speckled surface texture 
appears because the numerical simulation is trying to find equipotential 
lines in a random set of surface field variations very close to zero. 
This is the signature of a dark mode.

Similar dark modes appear in some other cavities with curved boundaries, 
but they all appear to be toroidal, independent of the topology of the cavity.
Toroidal cavities are best adapted to the geometry and topology of dark modes,
and they are also easy to fabricate and polish for minimising loss and thereby 
obtaining a better Q-value.

Notice that ${\rm TM}_{010}$ only has support on a smaller torus nested 
inside the toroidal cavity. 
Since the electric field of a dark mode vanishes everywhere on the boundary, 
it is potentially an ultrahigh-Q mode that perhaps can be used for long-term 
storage of quantum information. 
In short, since dark modes have minimal electric leakage from the cavity, 
and nodal modes have minimal magnetic leakage, the \emph{toroidal dark 
nodal mode} $\rm DNM = TM_{010} (\epsilon\rightarrow 1)$ is the lowest 
frequency candidate for an ultrahigh-Q mode.

\begin{figure}[b] 
\includegraphics[scale=0.13]{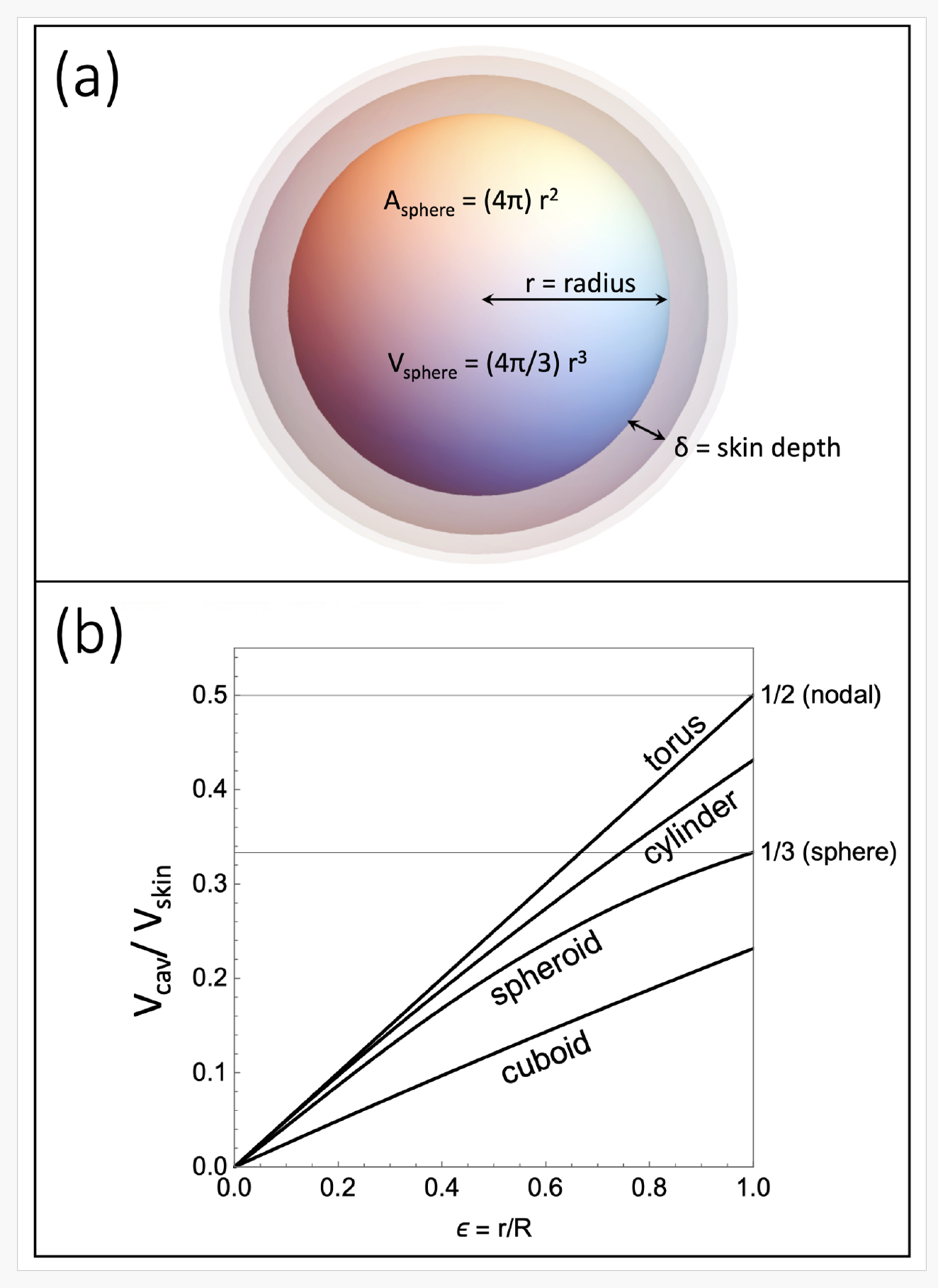}
\caption[Quality of geometry]
{Strategies for obtaining high and ultrahigh quality factors include: maximising 
the ratio of the cavity volume to its area, various types of surface treatment, 
using high purity aluminium and/or niobium, and storing quantum information in
``lossless" modes that minimise contact with the surface of the cavity.}
\label{fig:Qrat}
\end{figure}

\section{Quality of cavities}

There are many things that affect the internal quality factor (Q) of a cavity,
including cavity and mode geometry and topology, surface morphology
(roughness, defects, dislocations and damage), and many different material properties 
(chemistry, contamination, crystal structure, grain size and homogeneity, etc.).

Macroscopic electromagnetism in solids and interfaces between solids and 
empty space is a delicate long distance approximation of microscopic 
electromagnetism, which ultimately derives from the underlying quantum
electrodynamics. Fortunately, in the microwave regime it suffices to consider
approximate macroscopic effective parameters like surface impedance and 
resistance, residual surface resistance, skin depth, penetration depth, etc.
The wavelength of GHz microwaves is centimetres, which is much larger the 
the skin depth $\delta_s$ of a good lossy conductor (micrometers) and the 
London penetration depth $\delta_L$ of a superconductor (nanometers).

For a good conductor the surface resistance $R_s$ is the real part of the surface impedance $Z_s = R_s + i X_s$:
\begin{equation*}
R_s = \Re\left[Z_s\right] = \Re\left[\frac{1- i}{\sigma\delta}\right]
= \frac{1}{\sigma\delta} = \sqrt{\frac{\omega\mu}{2\sigma}}
\end{equation*}
At 1\,GHz the intrinsic impedance $\eta$ of a good conductor like aluminium
or copper is five orders of magnitude larger than the intrinsic impedance 
$\eta_0 = \sqrt{\varepsilon_0/\mu_0} = 377\,\Omega$ of the vacuum state.

The problem of how to improve the surface properties 
in order to reach the ultra-high Q-values, routinely obtained
in superconducting radio frequency (SRF) cavities used for particle acceleration 
in high energy physics, will not be discussed here. 
But note that the open shallow semi-doughnuts used 
in the fabrication of toroidal cavities (cf.\;Fig.\;\ref{fig:ProtoNodal}) should make 
many of the complicated surface processing methods (etching, annealing, electropolishing, welding, etc.) developed for SRFs easier, perhaps also 
giving more predictable and consistent results.

Investigating these modes systematically at cryogenic temperatures
has so far not been possible, so it must suffice for now to collect supporting 
evidence at room temperature, both from real data and numerical simulations.

\begin{figure}[] 
\begin{center}
\includegraphics[scale=0.21]{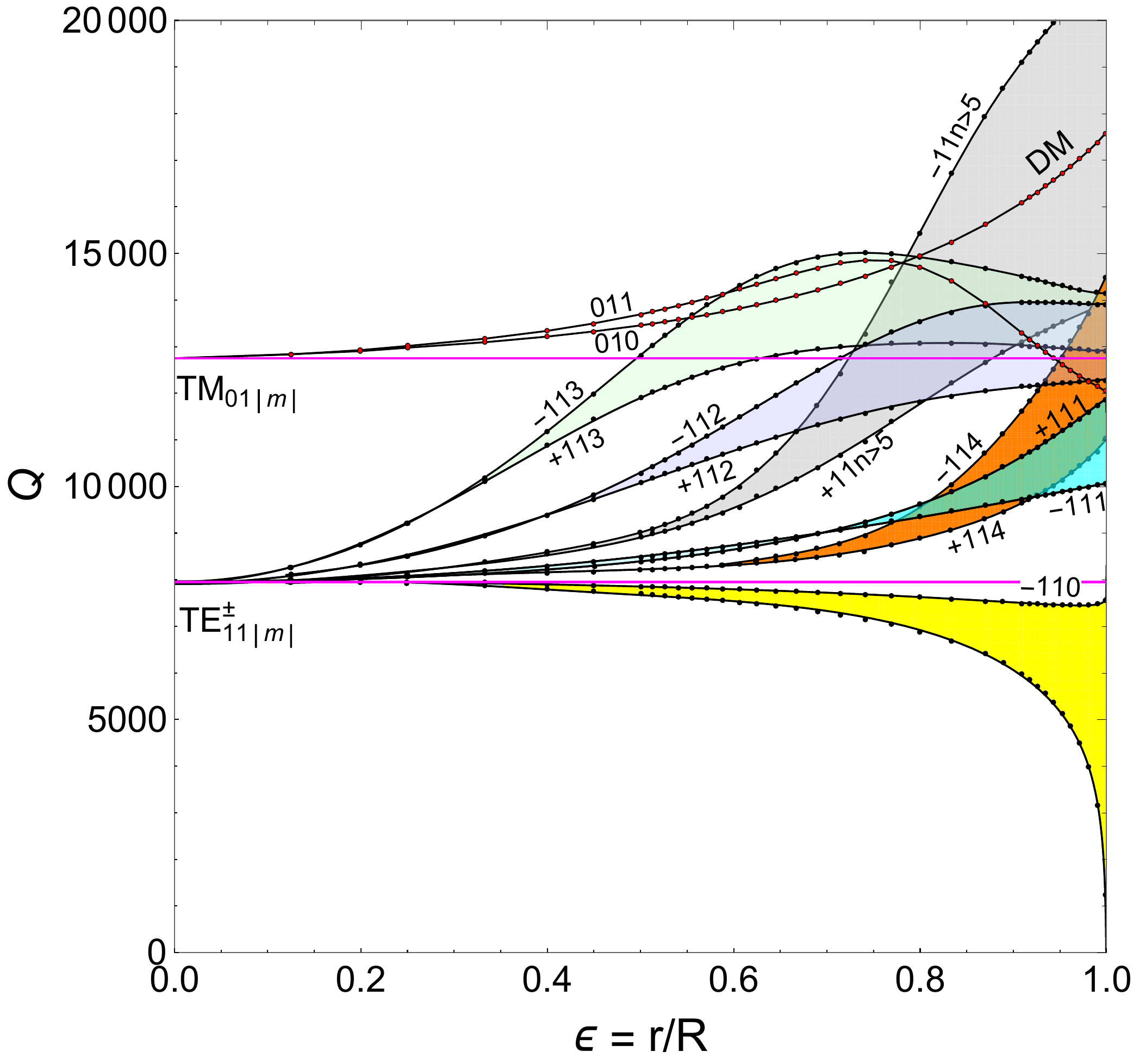}
\end{center}
\caption[Simulated toroidal Q-factors]
{Numerical simulation of the internal Q of some toroidal modes, 
as a function of the aspect ratio $\epsilon$.}
\label{fig:Qplot}
\end{figure}

Fig.\;\ref{fig:Qrat} shows a rough estimate of Q given by the dimensionless ratio 
$V_{cav}/(\delta A_{cav})$, where $A_{cav}$ ($V_{cav}$)
is the area (volume) of the cavity, and $\delta$ is the skin depth.
(a)\;The field at the surface of a lossy conductor penetrates a small depth 
$\delta$  into the cavity wall, where the resistance is 
exponentially damped but still finite. Resistive loss in the skin volume 
$V_{skin} = \delta A_{cav}$ degrades the internal quality of the cavity.
(b)\;Comparison of $V_{cav}/V_{skin}$ for four families of two-parameter cavities
of potential use in quantum processors. This estimate suggests that for 
$\epsilon \leq 1$ toroidal cavities have the highest Q-values, and that Q is maximal 
when the cavity is nodal ($r\rightarrow R$). Extensive numerical simulation has 
been used to verify that this is usually true, depending on the mode 
(cf.\;Fig.\;\ref{fig:Qplot}).

In a cylinder the lifetime $\tau_n$ of a photon occupying the mode with 
resonance frequency $f_n = c n/(2\pi \varepsilon_r R)$, 
in a medium with electrical permittivity
 $\varepsilon_r =\varepsilon/\varepsilon_0$ relative to the vacuum value 
 $\varepsilon_0$, is determined by the $Q$-value $Q = f_n/\Delta f_n$,
\begin{equation}
\tau_n = \frac{Q}{2\pi f_n} = \frac{\varepsilon_r}{c} \,\frac{Q R}{n},\quad n = 1,2,\dots,
\end{equation} 
where $c$ is the speed of light in the medium, and $\Delta f_n$ is the full width at
half maximum of the $n$'th resonance. Notice that the longest living mode 
in a cylinder is the ground state.

Inside an evacuated toroidal cavity with diameter $2R\sim 6\,cm$, 
$c = c_0 \approx 3\times 10^8\,m/s$ is the speed of light in vacuum 
and $\varepsilon_r = 1$, so the lifetime of a photon may be expected
to be roughly $\tau \sim Q\times 10^{-10}\, s$.
Given that $Q$-values exceeding $10^{11}$ now are routinely obtained in 
SRF cavities used in high energy particle accelerators \cite{RPZFABPG:2020}, 
a photon in a highly polished toroidal cavity inside a superconducting block of 
aluminium or niobium (the temperature in the cryostat is typically around
$10\,mK \ll T_c \sim 1\,K$) can live for several seconds. 
By that time transmon qubits, which typically have coherence times around
$10\,\mu s$, will have died a million times. 
The torus acts like a ``quantum bus" that entangles the qubits, as long as the 
photons are not absorbed by the cavity walls. 
How much information can be entangled in bosonic
quantum states (photons) in this cavity? In other words, can this serve 
as a quantum memory device?

\section{Quantum computing}

Cavities may be used for quantum computing by coupling the radiation field inside
the cavity (linear resonators) to the environment via non-linear resonators 
(artificial atoms, like transmons), for writing and reading information. 
The atomic state is not directly accessible, but since the atom hybridises 
with the photonic states we can interrogate the radiation field to infer 
what is going on. The atoms have very brief interactions with photons 
inside the cavity, typically 10\,-\,100\,$\mu$s, before the quantum information 
in the transmon is destroyed by decoherence. Since these states are not robust vigorous error correction is required to protect the quantum state.
Consequently, many, perhaps thousands, of physical qubits (circuit elements)
are required for each fully functional error corrected logical qubit used to build 
abstract logical gates. In short, this scheme is not hardware efficient.

However, a much less brute-force and hardware efficient alternative is available. 
We can exploit the huge Hilbert space of bosons (photons) to encode logical qubits
directly into the radiation field, by populating Fock states or other non-classical 
states \cite{Yale, NQ, NQ2}, without the need for much error correction.

The non-linear transmon is still required 
for brief interactions with the environment (reading and writing), for which its 
short lifetime is not a limiting factor. Since photons do not interact with each
other, the lifetime of these bosonic states is largely limited by properties of the 
cavity, including its geometry and topology, as discussed here. 
Virtues of tori include ease of manufacturing and polishing.

\begin{figure}[] 
\begin{center}
\includegraphics[scale=0.14]{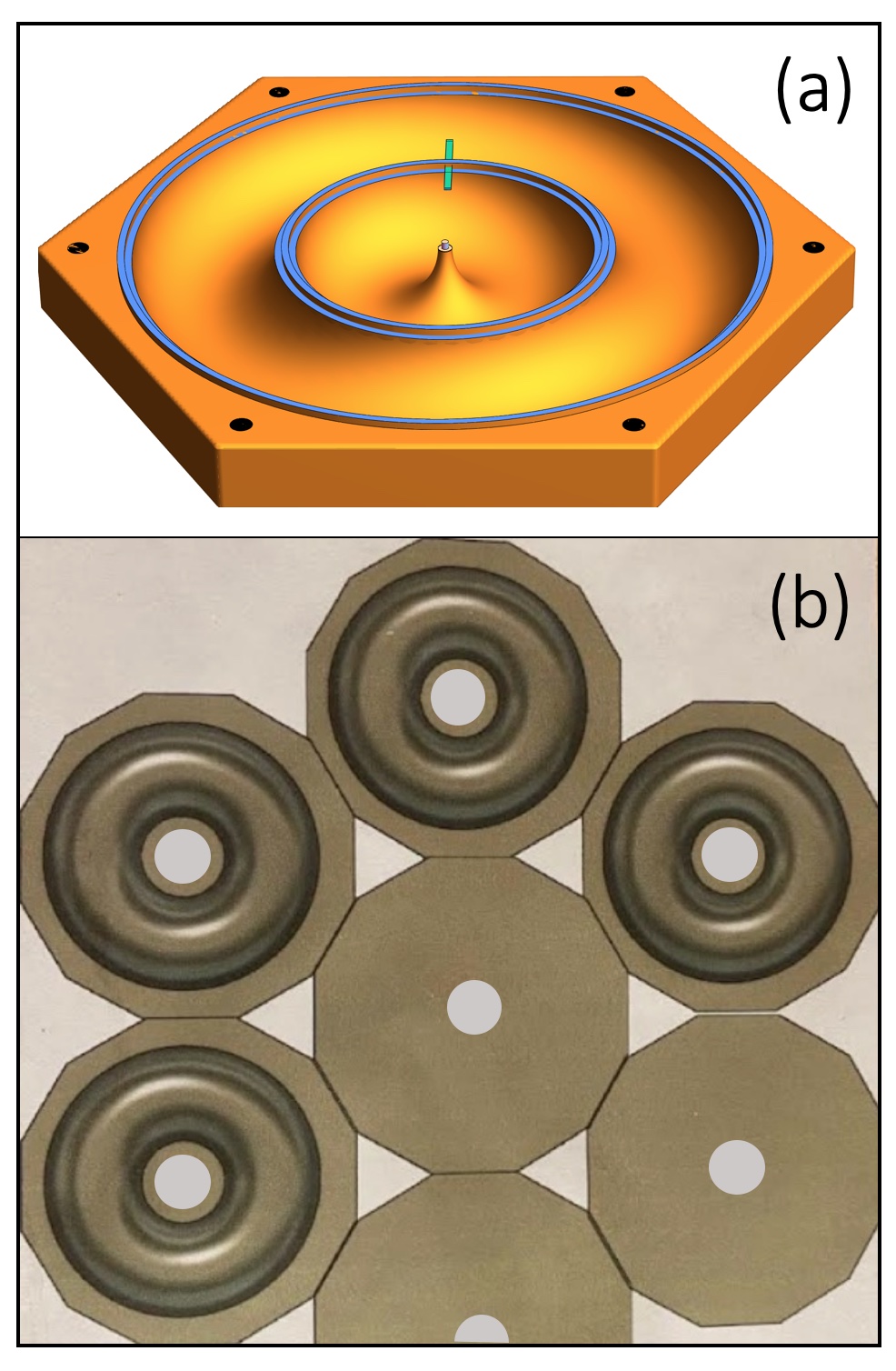}
\end{center}
\caption[DD]{(a) The inner nodal cavity in a double donut can be used 
for long time storage of quantum information.
(b) Honeycomb architecture allows scaling in two and three dimensions.}
\label{fig:DD}
\end{figure}

In an ultra-high-Q cavity photon decoherence may take several seconds, 
so minimal error correction is required, and it is also possible to store several 
non-interacting logical qubits directly in the bosonic modes, with little overhead. 

It may be advantageous to store information in an ultra-high-Q toroidal cavity 
that is only coupled through a single port to a readout torus, whose quality can 
be much lower [cf.\;Fig.\;\ref{fig:DD} (a)].

If we can scale up the number of 3D qubits, hundreds of logical qubits may
be packed into a relatively small volume that fits inside a standard cryostat.
One possibility is to stack the hexagonal cavity shown in Fig.\;\ref{fig:ProtoNodal} 
(right) in single- or multi-layered configurations [cf.\;Fig.\;\ref{fig:DD} (b)] 
(hexagons provide the densest packing in a plane). If only one or two layers
are used, then all cavities can be directly accessed from above or below.
 If nodal cavities are not used, then the donut holes provide access to all 
cavities in a multi-layered stack. If nodal cavities are used, then chopping 
corners off the hexagons makes dodecagons that leave triangular gaps for 
cabling to any cavity in a multi-layered device. In three dimensions it is 
possible to position the resonators to have direct straight connections 
(waveguides, say) between every pair of cavities.

A cavity stack can be manufactured from thin slabs ($\lesssim 2$ cm) of 
pure aluminium or niobium by machining semi-donuts on both sides of the slab
(except for the top and bottom layer), which are shallow and easy to polish.
After drilling ``port-holes" for cabling, stacks of these thin slabs should have good
thermal properties. If dark modes are used the orientation of the cavity seam does
not facilitate photon escape, providing some protection against decoherence.

A very rough estimate of the computing power of such a device is
obtained by considering 10\,-\,20\,GHz modes, for which the volume of each 
cavity is only a few cubic centimeters. We may be able to fit a few hundred 
of these cavities in a 100 litre cryostat. The number of logical qubits
could be in the thousands, but much less (20, say) would be a game-changer.

\section*{Acknowledgement}
I am grateful for assistance at the beginning of this project, including
discussions with Pol Forn-Diaz [Institut de F\' isica d'Altes Energies (IFAE),
Universidad Autonoma de Barcelona (UAB)], Ramiro Sagastizabal 
(Rigetti Computing), fabrication of the first toroidal cavity by Carlos Arteche [IFAE@UAB], and help with early COMSOL simulations by B. AlFakes 
[Technology Innovation Institute (TII)]. Discussions with Frederico Brito 
(TII), which led to the mode labelling used here and the improved VNA 
plot shown in Fig.\;\ref{fig:AluModeChart}, were especially helpful.

\bibliography{DesignTor_Biblio_2025-06-08.bib}

\end{document}